\documentclass{aa}
\include{epsf}

\begin{document}

\title{Magnetic activity in late-type giant stars: Numerical MHD simulations
 of non-linear dynamo action in Betelgeuse}

\author{S.B.F. Dorch }

\offprints{S.B.F. Dorch --- dorch@astro.ku.dk}

\institute{
     The Niels Bohr Institute for Astronomy, Physics and Geophysics,
     Juliane Maries Vej 30, DK-2100 Copenhagen {\O}, Denmark
}

\date{Received date, accepted date}

\authorrunning{Dorch}
\titlerunning{Magnetic activity in late-type giant stars}

\newcommand{\grad}    {\nabla}
\newcommand{\Div}     {\nabla\cdot}
\newcommand{\curl}    {\nabla\times}
\newcommand{\Laplace} {\nabla^2}
\newcommand{\rot}     {\curl}
\newcommand{\erfc}    {\mathop{\rm erfc}\nolimits}
\newcommand{\erf}     {\mathop{\rm erf}\nolimits}
\newcommand{\vekt}[1] {\mathbf{#1}}
\newcommand{\const}   {\mbox{\rm const}}
\newcommand{\Av}            {\vekt{A}}
\newcommand{\Bv}            {\vekt{B}}
\newcommand{\Fv}            {\vekt{F}}
\newcommand{\jv}            {\vekt{j}}
\newcommand{\uv}            {\vekt{u}}
\newcommand{\Heat}          {{\cal H}}
\newcommand{\Cool}          {{\cal C}}
\newcommand{\Strain}        {\bf {\mathsf{S}}}

\abstract{Evidence is presented from numerical
magneto-hydrodynamical simulations for the existence of magnetic
activity in late-type giant stars. A red supergiant with stellar
parameters similar to that of Betelgeuse ($\alpha$ Orionis) is
modeled as a ``star-in-a-box" with the high-order ``Pencil Code".
Both linear kinematic and non-linear saturated dynamo action are
found: the non-linear magnetic field saturates at a
super-equipartition value, while in the linear regime two
different modes of dynamo action are found. It is speculated that
magnetic activity of late-type giants may influence dust and wind
formation and possibly lead to the heating of the outer
atmospheres of these stars.  }

\maketitle

\section{Introduction}

There are indications from both dynamo theory and observations
that some late-type giant stars such as red supergiants and
asymptotic-giant-branch stars (AGB stars) may harbor magnetic
fields. On the theoretical side, it has been suggested that
non-spherically symmetric planetary nebulae (PNe) may be a result
of the collimating effect of a strong magnetic field: Blackman et
al.\ (\cite{Blackman+ea01}) studied interface dynamo models
similar to the mean field theory's solar $\alpha \omega$-dynamo
and found that the generated magnetic fields were strong enough to
shape bipolar outflows, producing bipolar PNe, while also braking
the stellar core thereby explaining the slow rotation of many
white dwarf stars. Also using mean field dynamo theory Soker \&
Zoabi (\cite{Soker+Zoabi2002}) propose instead an $\alpha^2
\omega$ dynamo due to the slow rotation of AGB stars rendering the
$\omega$-effect ineffective. They find that the magnetic field may
reach strengths of $\sim 100$ Gauss, significantly less than that
found by Blackman et al.\ (\cite{Blackman+ea01}). On the one hand,
they believe that the large-scale field is strong enough for the
formation of magnetic cool spots (see also Soker \& Kastner
\cite{Soker+Kastner2003} on AGB star flaring). These spots in turn
may regulate dust formation, and hence the mass-loss rate, but the
authors argue that they cannot explain the formation of
non-spherical PNe (see also Soker \cite{Soker2002}): on the other
hand, the locally strong magnetic tension could enforce a
coherent flow that may favor a maser process.

On the observational side of things, maser polarization is known
to exist in circumstellar envelopes of AGB stars (e.g.\ Gray et
al.\ \cite{Gray+ea99}, Vlemmings et al.\ \cite{Vlemmings+ea03},
and recently Sivagnanam \cite{Sivagnanam2004}) and X-ray emission
has been observed from some cool giant stars (e.g.\ H\"{u}nsch et
al.\ \cite{Hunch+ea98} and Ayres et al.\ \cite{Ayres+ea03}). These
observations are generally taken as evidence for the existence of
magnetic activity in late-type giant stars (cf.\ Soker \& Kastner
\cite{Soker+Kastner2003}).

The cool star Betelgeuse (a.k.a.\ $\alpha$ Orionis) is an example
of an abundantly observed late-type supergiant that displays
irregular brightness variations interpreted as large-scale surface
structures (e.g.\ Lim et al.\ \cite{Lim+ea98} and Gray
\cite{Gray2000}). It is one of the stars with the largest apparent
sizes on the sky---corresponding to a radius in the interval
600--800 ${\rm R}_{\odot}$. Freytag et al.\ (\cite{Freytag+ea02})
performed detailed numerical 3-d radiation-hydrodynamic (RHD)
simulations of the convective envelope of the star under realistic
physical assumptions, while trying to determine if the star's
known brightness fluctuations may be understood as convective
motions within the star's atmosphere: the resulting models were
largely successful in explaining the observations as a consequence
of giant-cell convection on the stellar surface, very dissimilar
to solar convection. Dorch \& Freytag (\cite{Dorch+Freytag2002})
performed a kinematic dynamo analysis of the convective motions
(i.e.\ not including the back-reaction of the Lorentz force on the
flow) and found that a weak seed magnetic field could indeed be
exponentially amplified by the giant-cell convection on a
time-scale of about 25 years.

This paper reports on full non-linear magneto-hydrodynamical (MHD)
numerical simulations of dynamo action in a late-type supergiant
star with fundamental stellar parameters set equal to that of
Betelgeuse. The paper is organized as follows: Section \ref{model}
contains a description of the numerical model, code and setup,
Section \ref{results} describes and discusses the results, the
convective flows and dynamo action, and Section \ref{conclusion}
contains a short summary and conclusion.

\section{Model}
\label{model}

The full 3-d MHD equations are solved for a fully convective star.
This is an example of a ``star-in-a-box'' simulation, where the
entire star is contained within the computational box. The
computer code used is the ``Pencil Code'' by Brandenburg \&
Dobler\footnote{{\tt
http://www.nordita.dk/data/brandenb/pencil-code/}}. The code has
been employed in several astrophysical contexts including e.g.\
hydromagnetic turbulence (see Brandenburg \& Dobler
\cite{Brandenburg+Dobler2002}, Dobler et al. \cite{Dobler+ea03}
and Haugen et al. \cite{Haugen+ea03}). The numerical method
performs well on many different computer architectures especially
on MPI machines, and uses 6th-order spatial derivatives and a
$2N$-type 3rd-order Runge-Kutta scheme. The code has a
``convective star'' module that allows the solution of the
non-linear MHD equations by the numerical pencil scheme in a star
with a fixed radius R and mass M. The code solves the following
general form of the compressible MHD equations:
\begin{eqnarray}
  \frac{{\rm D}\ln\varrho}{{\rm D}t}
  & = & - \Div\uv, \\
  \frac{{\rm D}\uv}{{\rm D}t}
  & = & -c_{\rm S}^2\grad\biggl(\frac{s}{c_p} + \ln\varrho\biggr)
      - \grad\Phi
      + \frac{\jv\times\Bv}{\varrho}  \nonumber \\
  & + & \nu \left( \Laplace\uv + \frac{1}{3}\grad\Div\uv
      + 2 \Strain \cdot\grad\ln\varrho\right), \\
  \frac{\partial\Av}{\partial t}
  & = & \uv\times\Bv - \eta\mu_0\jv, \\
  \varrho {\rm T}\frac{{\rm D}s}{{\rm D}t}
  & = &  \Heat - \Cool
      + \Div(K\grad {\rm T})
      + \eta\mu_0 \jv^2 + 2\varrho\nu\Strain^2,
\end{eqnarray}
where Eq.\ (1) is the mass continuity equation and Eq.\ (2) is the
equation of motion. Here $\varrho$ is density, $\uv$ the fluid
velocity, $t$ is time, ${\rm D}/{\rm D}t \equiv
\partial/\partial t + \uv\cdot\grad$ is the comoving derivative,
$c_{\rm S}$ is the sound speed, $s$ is the entropy, $\Phi$ is the
gravity potential, $\jv = \nabla\times \Bv/\mu_0$ the electric
current density, $\Bv$ the magnetic flux density, $\nu$ is
kinematic viscosity, and $\Strain$ is the traceless rate-of-strain
tensor.
Eq.\ (3) is the induction equation where $\Av$ is the magnetic
vector potential, $\Bv = \curl\Av$ the magnetic flux density,
$\eta = 1/(\mu_0\sigma)$ is the magnetic diffusivity ($\sigma$
being the electrical conductivity), and $\mu_0$ the magnetic
vacuum permeability. Eq.\ (4) is the energy, or rather entropy
equation, where T is temperature, $c_p$ the specific heat at
constant pressure, $\Heat$ and $\Cool$ are explicit heating and
cooling terms, and $K$ is the thermal conductivity (radiation is
not taken into account in MHD).

Variables are computed in terms of R and M so that e.g.\ the unit
of the star's luminosity L becomes $\frac{\rm M}{\rm R} ({\rm
GM}/{\rm R})^{\frac{3}{2}}$. In the present case these fundamental
parameters are set to ${\rm R} = 640~ {\rm R}_{\odot}$ and ${\rm
M} = 5~ {\rm M}_{\odot}$ yielding a luminosity of ${\rm L} =
46000~ {\rm L}_{\odot}$, consistent with current estimates of
Betelgeuse's size, mass and luminosity. The model employs a fixed
gravitational potential $\Phi$, an inner tiny heating core
entering into Eq.\ (4) through $\Heat$, and an outer thin
isothermally cooling spherical surface at $r = {\rm R}$ (corresponding to
$\Cool$ in Eq.\ 4), with a Newtonian cooling time scale set to
$\tau_{\rm cool} = 1$ year corresponding to the typical convective
turn-over time in the model of Freytag et al.
(\cite{Freytag+ea02}).

Due to constraints on computer time and resolution the true
thermodynamic range of the star can not be represented and the
surface is cooled at a temperature that is 4.5 times higher than
Betelgeuse's effective temperature which is about T$_{\rm eff} =
3500$ K (various estimates exist in the literature, see e.g.\
Freytag et al. \cite{Freytag+ea02}). The fixed gravity is
approximately given by a 1/r-potential (as in Freytag et al.\
\cite{Freytag+ea02}) and initially the star expands resulting in a
slight decrease of 1.4\% of the mean mass density, subsequently it
re-contracts by roughly 0.1\%. Betelgeuse is only slowly rotating
and a rotational frequency was chosen corresponding to a surface
rotational velocity of 5 km/s, yielding a large Rossby number.

Models with different numerical resolutions have been run for
testing: in this paper the results come from a model with $128^3$
uniformly distributed grid points yielding a spatial resolution of
$\Delta{\rm x} = 15 {\rm R}_{\odot}$ the physical size of the box
being R$^3$. The models were computed at the Danish Center for
Scientific Computing Horseshoe 512 cpu Linux cluster typically
allocating between 16 and 32 cpus.

Boundary conditions on the computational box are anti-symmetric
for components of the vector fields $\uv$ and $\Bv$ in the
direction of the component (yielding a vanishing value at the
boundary) and symmetric across the boundary in the direction
perpendicular to the component (yielding a vanishing gradient
across the boundary). Boundary conditions for the density $\ln
\rho$ are anti-symmetric with respect to an arbitrary value across
all boundaries (yielding a vanishing second derivative), and the
boundary condition for the entropy $s/c_p$ corresponds to a
constant temperature at the boundary.

\subsection{Magnetic Reynolds number}

Dynamo action by flows are often studied in the limit of
increasingly large magnetic Reynolds numbers Re$_{\rm m} = \ell
{\rm U}/\eta$, where $\ell$ and U are characteristic length and
velocity scales. Most astrophysical systems are highly conducting
yielding small magnetic diffusivities $\eta$, and their dimensions
are huge resulting in huge values of Re$_{\rm m}$. Betelgeuse is
not an exception and most parts of the star is better conducting
than the solar photosphere that has a magnetic diffusivity of the
order of $\eta \approx 10^4$ m$^2$/s:

\begin{figure}[!htb]
\makebox[8cm]{ \epsfxsize=8.0cm \epsfysize=6.0cm \epsfbox{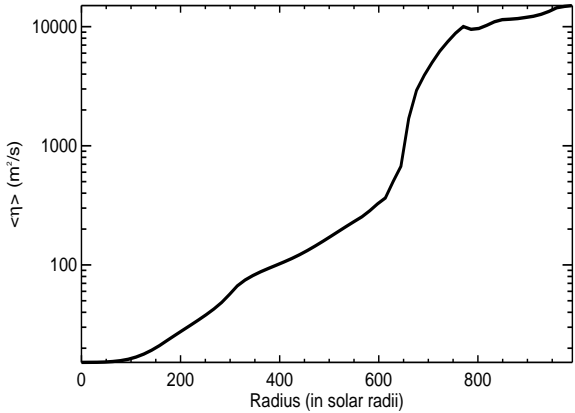}
} \caption[]{Diffusivity: average radial magnetic diffusivity
$\eta$ (m$^2$/s) in the model of Betelgeuse as a function of
radial distance in solar radii R$_{\odot}$. Adopted from Dorch \&
Freytag \cite{Dorch+Freytag2002}.} \label{fig1}
\end{figure}
Figure \ref{fig1} shows the average Spitzer's resistivity as a
function of radius in the model of Betelgeuse by Freytag et al.\
(\cite{Freytag+ea02}). Spitzer's formula (e.g.\ Schrijver \& Zwaan
\cite{Schrijver+Zwaan2000}) assumes complete ionization and hence
the precise values of $\eta$ are uncertain in the outer parts of
the star, where the atmosphere borders on neutral. There is some
uncertainty connected also with defining the most important length
scale of the system, but preliminarily taking $\ell$ to be 10\% of
the radial distance R from the center (a typical scale of the
giant cells), and U $= {\rm u}_{\rm RMS}$ along the radial
direction yields Re$_{\rm m}=10^{10}$--$10^{12}$ in the interior
part of the star where R $\le 700$ R$_{\odot}$.

In the present case we cannot use that large values of Re$_{\rm
m}$ (partly due to the fact that we are employing a uniform fixed
diffusivity $\eta$), but rely on the results from generic dynamo
simulations indicating that results converge already at Reynolds
numbers of a few hundred (e.g.\ Archontis, Dorch, \& Nordlund
\cite{Archontis+ea03a,Archontis+ea03b}). Furthermore, Dorch \&
Freytag (\cite{Dorch+Freytag2002}) obtained kinematic dynamo
action in their model of a magnetic Betelgeuse at Re$_{\rm m} \sim
500$. Here we use an $\eta$ that is about $10^8$ times too large
compared to the estimated surface value in Betelgeuse, leading to
a magnetic Reynolds number of Re$_{\rm m} \sim 300$ (based on the
largest scales).

\section{Results and discussion}
\label{results}

There is some disagreement as to what one should require for a
system to be a ``true" astrophysical dynamo. Several ingredients
seem to be necessary: the flows must stretch, twist and fold the
magnetic field lines (e.g.\ Childress \& Gilbert \cite{STF});
reconnection must take place to render the processes irreversible;
weak magnetic field must be circulated to the locations where flow
can do work upon it (cf.\ Dorch \cite{Dorch2000}); and finally,
the total volume magnetic energy E$_{\rm mag} = \int_{\rm V} {\rm
e}_{\rm m} {\rm dV}$ must increase (the linear regime) or remain at
a constant saturation amplitude on a long time scale (non-linear
regime). These points are based largely on experience from
idealized kinematic and non-linear MHD dynamo models; e.g.\ Archontis
et al. (\cite{Archontis+ea03a,Archontis+ea03b}). This paper
deals mainly with the question of the exponential growth and
saturation of E$_{\rm mag}$.

\begin{figure}[!htb]
\makebox[9.0cm]{ \epsfxsize=9.0cm \epsfysize=9.0cm
\epsfbox{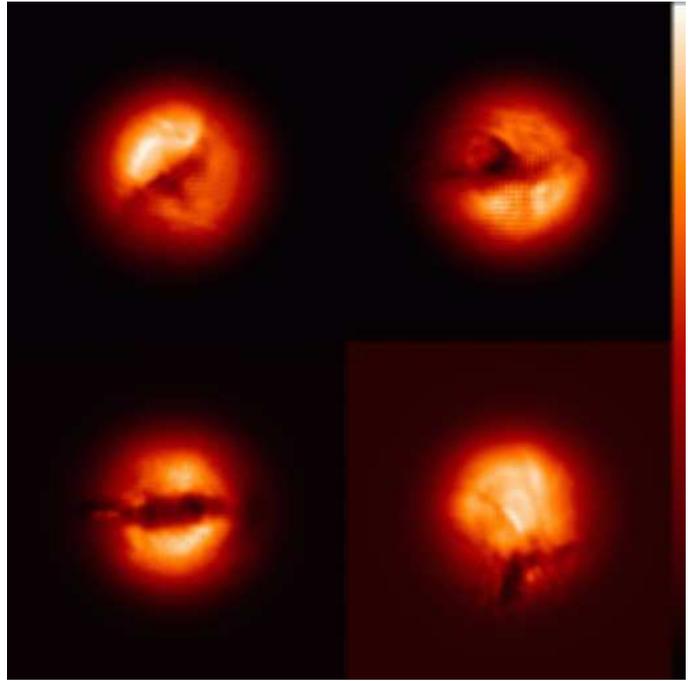} } \caption[]{Simulated surface
intensity snapshots at four different instants, time = 256, 347,
457 and 494 years (from upper left to lower right).} \label{fig2}
\end{figure}

\subsection{Convective flows}

Although not the main topic of this paper, it is in its place to
discuss here also the properties of the convective flows in the
model, since these supply the kinetic energy forming the basic
energy reservoir for any dynamo action that might be present. It
is not expected that the flows match exactly what is found in more
realistic RHD simulations, but at least a qualitative agreement
should be inferred since the fundamental parameters of this MHD
model and the RHD model of Freytag et al.\ (\cite{Freytag+ea02})
are the same.

The velocity is initialized with a random flow with a small
amplitude. Rapidly large-scale convection cells develop
through-out the star: the giant cell convection is evident in both
the thermodynamic variables, such as temperature and gas pressure, as
well as in the flow field, but the observational equivalent
however, is the surface intensity. Since the model does not
incorporate realistic radiative transfer (as opposed to the model
of Freytag et al.\ \cite{Freytag+ea02}), only a simulated
intensity can be derived. Using a frequency independent LTE source
function a simulated intensity I$_{\rm sim}$ can be defined as:
\begin{eqnarray}
 {\rm I}_{\rm sim}{\rm (y,z)} & = & \int_0^{\rm R} \sigma {\rm T(x,y,z)}^4 e^{-\tau{\rm (x,y,z)}} {\rm d}\tau{\rm (x,y,z)},\\
 \tau({\rm x}) & = & \int_0^{\rm x} \kappa_0 \rho{\rm (x,y,z)} {\rm dx},
\end{eqnarray}
where T is temperature, $\tau$ is a measure of optical depth along
one of the axes (here the radial {\em line of sight} direction is
taken to be the grid x-axis), $\rho$ is mass density, $\kappa_0$
is a constant gray opacity and $\sigma$ Stefan's constant.
Figure \ref{fig2} shows simulated intensity snapshots at four
different instances: the typical contrast between bright and dark
patches on the surface is 20--50\%, and only 2--4 large cells are
seen at the stellar disk at any one time corresponding to a hand
full of cells covering the entire surface. This is in qualitative
agreement with the RHD models of Freytag et al.\
(\cite{Freytag+ea02}): the surface is not composed of simply
bright granules and dark intergranular lanes in the solar sense---sometimes
the pattern is even the reverse of this---e.g.\ in Figure
\ref{fig2} the simulated intensity snapshot at time $t = 457$
years, the cool dark area in the center of the stellar disk is
actually a region containing an upward flow.

\begin{figure}[!htb]
\makebox[8cm]{ \epsfxsize=8.0cm \epsfysize=6.0cm
\epsfbox{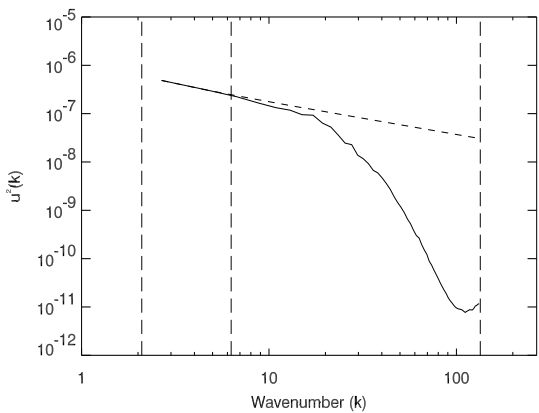} } \caption[]{Powerspectrum of ${\rm
u}^2(k)$ at time = 457 years, and a line corresponding to
Kolmogorov scaling (dashed line). The vertical lines, from left to
right, denote the wavenumbers corresponding to the computational
box size $k_0$, the stellar radius $k_{\rm R}$, and the Nyquist
frequency $k_{\rm Ny}$ (the numerical resolution). } \label{fig3}
\end{figure}
More quantitatively, the kinetic power-spectrum (Fig.\
\ref{fig3}) illustrates that there is much more power on large
scales than on the small scales of the velocity field: below a
wavenumber of $k \sim$ 20 (based on the box size in units of the
star's radius) power is decreasing fast, but at larger scales the
power is proportional to $k^{-2/3}$ corresponding to normal
Kolmogorov scaling, the inertial range spans however only roughly
one order of magnitude. In conclusion the large-scale convective
patterns are then typically larger than 15--30\% of the radius,
and are actually often on the order of the radius in size. The
corresponding radial velocities range between 1--10 km/s in both
up and down flowing regions.

There are at least three different evolutionary phases of
convection in the simulations, depending on the level of the total
kinetic energy E$_{\rm kin}$ of the convection motions: initially
there is a transient of about 30 years after which the RMS
velocity field reaches a level where it fluctuates around a value
of about 800 m/s (this corresponds to the kinematic phase of the
dynamo, where the flow is unaffected by the presence of the still
weak magnetic field, see below). During the rest of the simulation
after about 290 years, the RMS speed measured in the entire box
decreases to 500 m/s (when
the energy in the magnetic field becomes comparable to the kinetic
energy density). During the entire simulation, however, the
maximum speed in the computational box fluctuates around a
constant value of about 90 km/s. The flows are not particularly
helical and the mean kinetic helicity is on the order of $10^{-6}$
m/s$^2$. Mean field $\alpha\omega$-type solar dynamos do not
produce large-scale fields if the kinetic helicity is less than a
certain value (cf.\ Maron \& Blackman \cite{Maron+Blackman2002})
and hence we cannot expect a large-scale toroidal field in the
solar sense to be generated.

\begin{figure}[!htb]
\makebox[8cm]{ \epsfxsize=8.0cm \epsfysize=6.0cm
\epsfbox{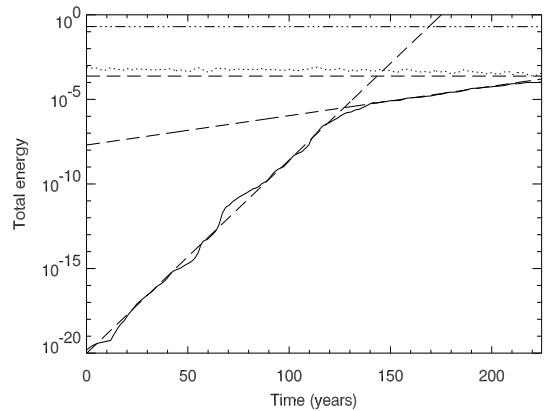} } \caption[]{Linear regime: energy as a
function of Betelgeusian time in years. The upper almost
horizontal line is total thermal energy E$_{\rm th}$
(dashed-dotted), middle curve with wiggles is total kinetic energy
E$_{\rm kin}$ (dotted), and lower full curve is E$_{\rm mag}$. The
three thin dashed lines corresponding to exponential growth with
characteristic time-scales of 3.8, 25 and $\infty$ years. }
\label{fig4}
\end{figure}

\begin{figure}[!htb]
\makebox[8cm]{ \epsfxsize=8.0cm \epsfysize=6.0cm
\epsfbox{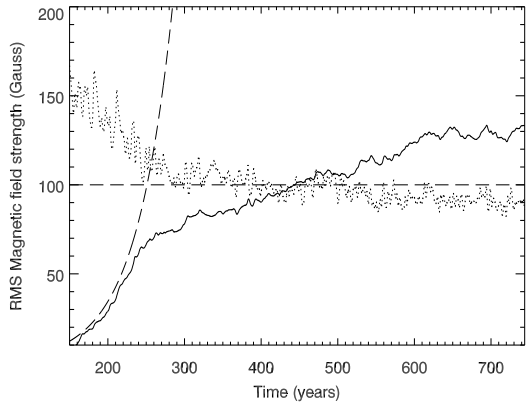} } \caption[]{Transition to the
non-linear regime: RMS magnetic field B in the whole computational
box in Gauss as a function of Betelgeusian time in years. The
upper curve is the equipartition field strength B$_{\rm eq}$
corresponding to the average kinetic energy density of the fluid
motions (dotted curve) and the lower full curve is the actual RMS
field strength B (full curve). The dashed thin curve correspond to
growth times of 25 years and a horizontal reference line at a
field strength of 100 Gauss. } \label{fig5}
\end{figure}

\subsection{Dynamo action}

In an earlier kinematic study of Betelgeuse using a completely
different numerical approach (Freytag et al.\ \cite{Freytag+ea02}
and Dorch \& Freytag \cite{Dorch+Freytag2002}), dynamo action was
obtained when the specified minimum value of Re$_{\rm m}$ was
larger than approximately 500 and at lower values of Re$_{\rm m}$
the total magnetic energy decayed. In the present case Re$_{\rm
m}$ is of the same order of magnitude and we find an initial clear
exponential growth over several turn-over times, and many orders
of magnitude in energy. Figure \ref{fig4} shows the evolution of
E$_{\rm mag}$ as a function of time, for the first 225 years (in
Betelgeusian time): initially there is a short transient, where
the field exponentially decays because the fluid motions has not
yet attained their final amplitude. Once the giant cell convection
has properly begun however, the magnetic field is amplified and we
enter a linear regime of exponential growth. There are two modes
of amplification in the linear regime; the initial mode with a
growth rate of about 4 years, which in the end gives way to a mode
with a smaller growth rate corresponding to 25 years. This is a
slightly unusual situation, since normally modes with smaller
growth rates are overtaken by modes with larger growth rates (cf.\
Dorch 2000); the explanation is that while both modes are growing
modes, only the one with the largest growth rate is a purely
kinematic mode---while the exponential growth of the second mode
is linear it is not kinematic---the presence of the magnetic field
is felt by the fluid through the back-reaction of the Lorentz
force in Eq.\ (2) becoming important. This quenches the growth
slightly and henceforth one can refer to the second mode as a
``pseudo-linear" mode.

No exponential growth can go on forever and eventually the
magnetic energy amplification must come to a halt: the question is
then whether the magnetic field retains a more or less constant
saturation value, or if it dissipates. The latter is only possible
if the non-linear mode is a decaying mode that corresponds to a
negative growth rate. In case of saturation the typical field
strength is expected to be on the order of the equipartition value
corresponding to equal magnetic and kinetic energy densities.
Figure \ref{fig5} shows the RMS magnetic field strength B$_{\rm
RMS}$ within the entire model star as a function of time for $\sim
700$ Betelgeusian years: the pseudo-linear mode as well as the
mode in the non-linear regime are shown. The RMS magnetic field
saturates at a value slightly above the RMS equipartition field strength
B$_{\rm eq} = \sqrt{\mu_0 \left< \rho {\rm u}^2 \right>}\sim$ 90--100 Gauss,
corresponding to a value of about 120--130 Gauss. In terms of total
energy this means that the magnetic energy E$_{\rm mag}$ is above
equipartition with the kinetic energy E$_{\rm kin}$ by roughly a
factor of two.
Hence the field cannot be said to be extremely strong, but it is
not particularly weak in most parts of the star either. What may be
interesting from an observational point of view is the strength of
the field at the surface. In the non-linear regime the field
strength at the sphere with radius $r = {\rm R}$ can be up to
$\sim 500$ Gauss, while in the interior of the star it can be as high
as a few kG:
the strong field of the intermittent magnetic structures almost completely
quenches the velocity field in these regions that are small-scale
compared to the scale of the convection; i.e.\ the local field can be far
above equipartition.

\begin{figure}[!htb]
\makebox[8cm]{ \epsfxsize=8.0cm \epsfysize=6.0cm
\epsfbox{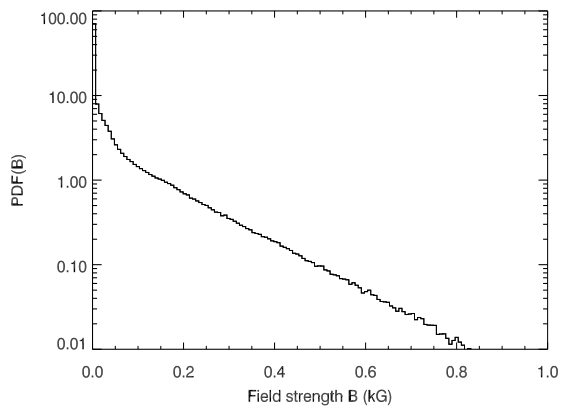} } \caption[]{Distribution of magnetic field
strength: PDF for the magnetic field $|{\rm B}|$ (in kG) at time =
732 years within the non-linear saturated regime.} \label{fig6}
\end{figure}

\begin{figure}[!htb]
\makebox[8cm]{ \epsfxsize=8.0cm \epsfysize=6.0cm
\epsfbox{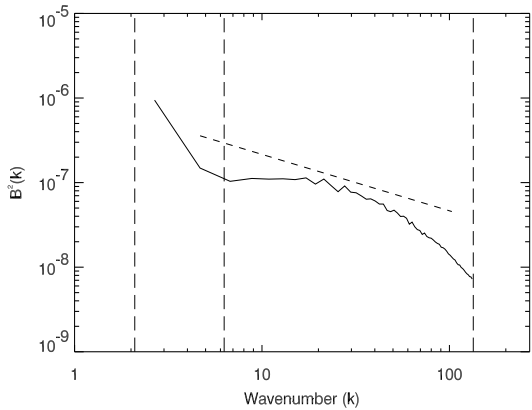} } \caption[]{Powerspectrum of the magnetic
field B$(k)^2$ (solid curve) at time = 732 years, and a line
corresponding to a power-law with an exponent of -2/3 (dashed
line). The vertical lines, from left to right, denote the
wavenumbers corresponding to the computational box size $k_0$, the
stellar radius $k_{\rm R}$, and the Nyquist frequency $k_{\rm Ny}$
(the numerical resolution). } \label{fig7}
\end{figure}

\begin{figure*}[!htb]
\makebox[18.0cm]{ \epsfxsize=9cm \epsfysize=8cm
\epsfbox{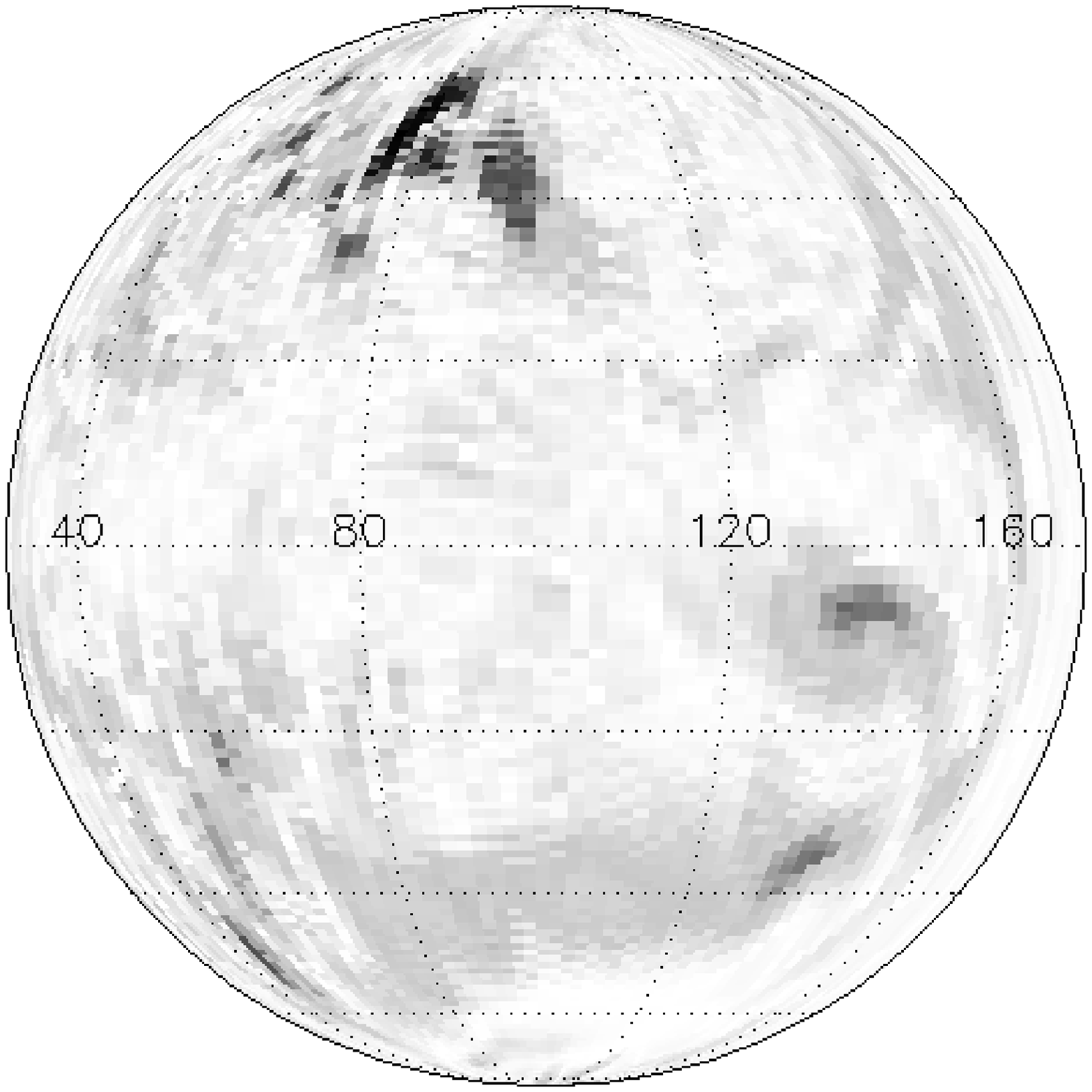} \epsfxsize=9cm \epsfysize=8cm
\epsfbox{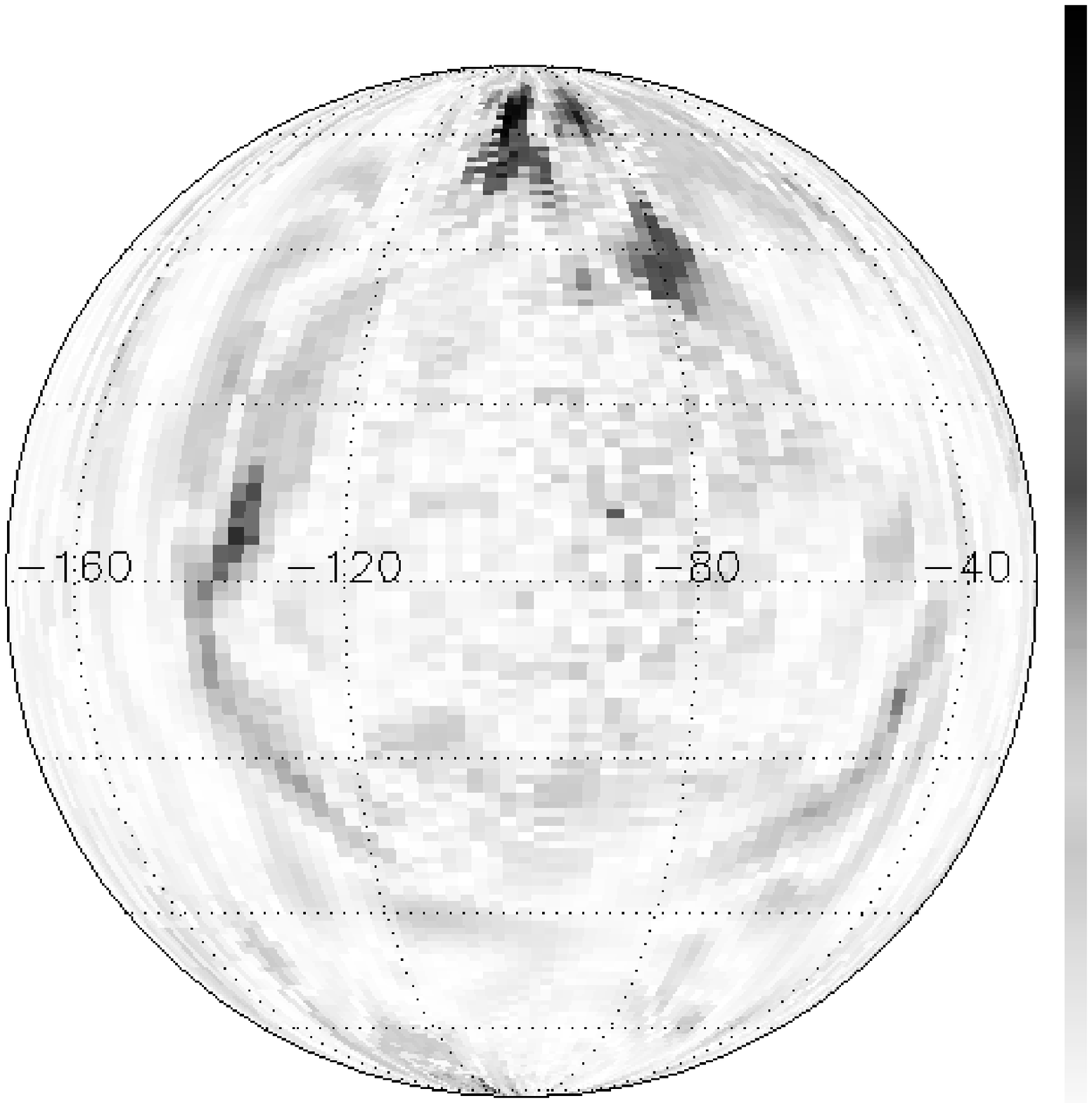} } \caption[]{An illustration of the
unsigned magnetic field strength $|{\rm B}|$ at the spherical
surface $r =$ R of the model star using an orthographic map
projection. The darkest patches correspond to a maximum field
strength of 500 Gauss (black on the continuous scale bar). From a
snapshot at time = 695 years. The views are centered on longitudes
of $100^{\rm o}$ (left) and $-100^{\rm o}$ (right). The grid indicated has a 
longitudinal spacing of $40^{\rm o}$ and a latitudinal spacing of $20^{\rm o}$. 
The numerical resolution of the map is $180^2$ grid points. } 
\label{fig8}
\end{figure*}

\begin{figure}[!htb]
\makebox[8.75cm]{ \epsfxsize=8.75cm \epsfysize=8cm
\epsfbox{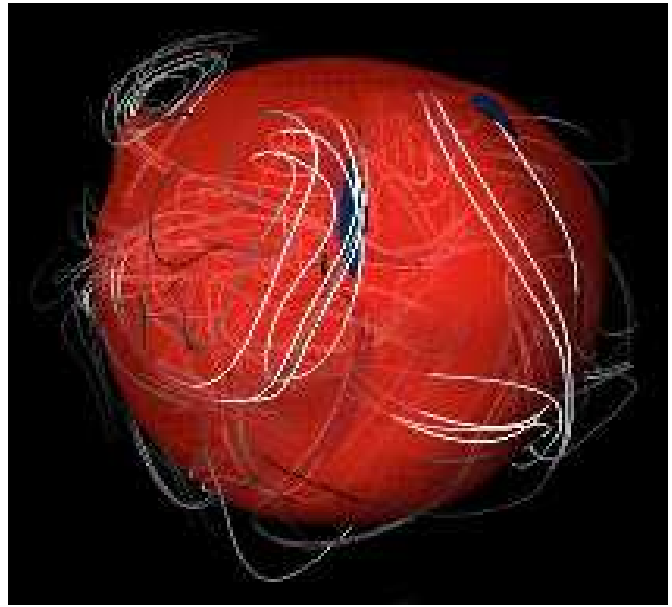} } \caption[]{A 3-d volume rendering of
magnetic field lines (white) and flux ropes (dark structures).
Also show is a isosurface (transparent) at a the surface
temperature value.} \label{fig9}
\end{figure}

\subsection{Magnetic structures}

It is interesting to examine the geometry of the magnetic field
that the saturating non-linear dynamo generates since this could
be relevant for the influence of the field on e.g.\ asymmetric
dust and wind formation. Qualitatively speaking the field becomes
concentrated into elongated structures much thinner than the scale
of the giant convection cells, but perhaps due to the very
irregular nature of the convective flows, no ``intergranular
network'' is formed in the solar sense. On the one hand, at times magnetic
structures coincide with downflows, but not as a general rule. On the
other hand, strong fields are seldomly located within the general
upflow regions.

Figure \ref{fig6} shows the PDF of the magnetic field: the
distribution is a typical signature of highly intermittent
structures, i.e.\ only a very small fraction of the volume carries
the strongest structures and the probability of finding a
vanishing field strength at a random point in space, is far
greater than finding strong fields.

An energy spectrum ${\rm B}^2(k)$ reveals that the magnetic
structures are well resolved with little power at the Nyquist
wavenumber $k_{\rm Ny} = \pi/\Delta{\rm x}$, and that the power at
the largest wavenumbers $k\sim100$ is two-orders of magnitude
smaller than that at the largest scales (see Figure \ref{fig7}).
Maximum power is obtained on the largest scales corresponding to
wavenumbers of a few, while there is a dip at $k\sim7$
corresponding to the scale of the radius, where the power is
minimum. The power on scales $k\approx$ 10--20 is flat leaning
towards being proportional to $k^{-2/3}$ corresponding to
Kolmogorov scaling, and at small-scales $k\la50$ power steeply
drops: magnetic structures in the non-linear regime are then large
by solar standards, but smaller than the giant convection cells
that show increasing power towards large scales.

Figure \ref{fig8} is a map of the spherical surface at $r =$ R in
terms of magnetic energy in the non-linear regime: there are both
bright patches of strong magnetic field, e.g.\ a 500 Gauss region
at longitude around $80^{\rm o}$ and $50^{\rm o}$ north of the equator, and
large areas with a vanishing field (e.g.\ at longitude $80^{\rm
o}$ on the equator). On this map the areal magnetic filling factor
is $\sim$55\% for $|{\rm B}|>50$ Gauss, while it is only
$\sim$0.6\% for $|{\rm B}|>500$ Gauss. 
However, this mapping does not represent
any physical surface of the star. Due to the fact that the actual
upper boundary consists of a few large cells the surface cannot be
captured by a simple sphere with radius R; this is illustrated by
a volume rendering (Fig.\ \ref{fig9}) of the isosurface at the
cooling temperature. E.g.\ in the upper left corner of Figure
\ref{fig9} there is a ``hill'' in the temperature isosurface and
further large ``slopes'' can be seen across the star in this
illustration. In the latter figure, the magnetic field lines
illustrate that there is a slight trend inside the star, looking
through the partly transparent surface, towards a radial
orientation of the magnetic field, while the strong fields near
the surface of the star are predominantly horizontally aligned.
This was also observed by Dorch \& Freytag
(\cite{Dorch+Freytag2002}) and may be a generic trade of
giant-cell full-radius convection in slowly rotating stars.

\section{Summary and conclusion}
\label{conclusion}

In summary three different modes of dynamo action are recognized:
\begin{enumerate}
 \item A relatively fast growing linear mode with an exponential
 growth time of $\sim 4$ years.
 \item A relatively slowly growing pseudo-linear mode with an exponential growth of
  $\sim 25$ years.
 \item A saturated non-linear mode operating a factor of two above
  equipartition.
\end{enumerate}
More modes may of course exist but these must then have very low
growth rates and/or very small initial amplitudes since they have
not appeared in the simulations. It is worth noting that in case
2) of the pseudo-linear mode, the same value of the growth time
(around 25 years) was found in the previous purely kinematic
dynamo models Dorch \& Freytag (\cite{Dorch+Freytag2002}) although
they employed a different computational method. This may in fact
not be so strange, since the growth rate in a kinematic dynamo is
set in part by the convergence of the flow across the field lines
$- \nabla_{\perp}\cdot\uv$ and if the flows are similar so should
the growth rates be.

Based on the results presented here, it is not possible to state
conclusively if Betelgeuse actually has a magnetic field, since
such a field is unobserved. However, one may conclude that it
seems possible that late-type giant stars such as Betelgeuse can
indeed have presently undetected magnetic fields. These magnetic
fields are likely to be close to or stronger than equipartition; they
may be difficult to detect directly, due to the relatively small
filling factors of the strong fields, but even the moderately strong fields
may have influence on
their immediate surroundings through altered dust, wind and
mass-loss properties. The formation of dust in the presence of a
magnetic field will be the subject of a subsequent paper along the
lines presented here: the ``Pencil Code'' has recently been augmented
with modules for radiation and dust modeling.

The dynamos of the late-type giant studied here may be characterized as
belonging to the class called ``local small-scale dynamos''
another example of which is the proposed dynamo action in the
solar photosphere that is sometimes claimed to be responsible for
the formation of small-scale flux tubes (cf.\ Cattaneo
\cite{Cattaneo1999}). However, in the case of Betelgeuse this
designation is less meaningful since the generated magnetic field
is both global and large-scale, but because of the slow
rotation, no large-scale solar-like toroidal field is formed.

It is interesting to note that very recently Lobel et al.\
(\cite{Lobel+ea04}) published spatially resolved spectra of the
upper chromosphere and dust envelope of Betelgeuse. Based on
various emission lines they provide evidence for the presence of
warm chromospheric plasma away from the star at around 40 R. The
spectra reveal that Betelgeuse's upper chromosphere extends far
beyond the circumstellar envelope. They compute that temperatures
of the warm chromospheric gas exceed 2600 K. The presence of a hot
chromosphere lead this author to speculate on the possible connection
to coronal heating in the Sun, which is likely to be magnetic in
origin and caused by flux braiding motions in the solar
photosphere (cf.\ Gudiksen \& Nordlund
\cite{Gudiksen+Nordlund2002}): it remains to be proven whether a
similar process could be operating in late-type giant stars.

\begin{acknowledgements}
This work was supported by the Danish Natural Science Research
Council. Access to computational resources granted by the Danish
Center for Scientific Computing in particular the Horseshoe
cluster at Odense University.
\end{acknowledgements}

\end{document}